\documentclass[preprint,12pt]{elsarticle}
\usepackage{amsmath}
\usepackage{amsfonts}
\usepackage{graphicx}
\usepackage{color}
\begin{document}

\title{The role of  elastic and inelastic processes in
the temperature dependence of Hall induced resistance oscillations
in strong magnetic fields.}
\author{Alejandro Kunold}
\ead{akb@correo.azc.uam.mx}
\address{Departamento de Ciencias B\'asicas,
Universidad Aut\'onoma Metropolitana-Azcapotzalco,
Av. San Pablo 180,  M\'exico D. F. 02200, M\'exico}

\author{Manuel Torres}
\ead{torres@fisica.unam.mx} 
\address{Instituto de F\'{\i}sica,
Universidad Nacional Aut\'onoma de M\'exico,
Apartado Postal 20-364,  M\'exico Distrito Federal 01000, M\'exico}

\date{\today}
%\pacs{73.43.Qt,71.70.Di,73.43.Cd,73.50.Bk,73.50.Fq}
\begin{abstract}

We develop a model of magnetoresistance oscillations induced by the Hall field
in order to study the    temperature dependence observed in recent
experiments.
The model is based on the solution of the von Neumann equation incorporating
the exact dynamics of two-dimensional damped electrons in the
presence of arbitrarily strong  magnetic  and  dc electric fields,
while the effects of randomly distributed neutral and charged impurities are
perturbatively added.
Both the effects of elastic impurity scattering
as well as those related to inelastic processes play an important role.
The theoretical predictions  correctly reproduce the
main  experimental features  provided that the inelastic scattering rate obeys a
$T^2$ temperature dependence, consistent with electron-electron
interaction effects.
 
\end{abstract}

\maketitle

\section{Introduction}

Magnetoresistance oscillations of two-dimensional electron systems (2DES), besides that Shubnikov-de Haas oscillations,  have been a considerably active 
topic of research over the past decade.
Microwave-induced resistance oscillations (MIRO) were
registered \cite{zudov:201311,zudov:046807,mani:646,ye:2193}
in high mobility semiconductor 2DES samples   subjected  to
microwave radiation and low
magnetic fields.
Similarly Hall field-induced resistance oscillations (HIRO) are observed  in   high
mobility samples in the presence of  an strong longitudinal dc-current excitation
\cite{yang:076801,bykov:245307, zhang:081305,zhang:041304}.
Even though the mechanisms that produce MIRO and HIRO are different, 
both   display periodic oscillations  in $1/B$  that originates in the
commensurability of the cyclotron frequency $\omega_c$ with a characteristic
parameter of the system. Remarkably,  in sufficiently clean samples the MIRO AND HIRO 
evolve into 
zero resistance or   zero  differential resistance states, respectively  \cite{bykov:116801,zhang:036805}.
Some models have been proposed to explain the main experimental features
of HIRO. On one hand the \emph{displacement model} relays on the impurity induced electron transitions
between  Landau levels (modified by the effects of the microwave radiation or the strong dc-excitation)
\cite{zhang:041304,lei:132119,kunold:3803}
on the other, the \emph{inelastic model} depends on the formation of a
non-equilibrium distribution function induced by
the Hall field \cite{vavilov:115331}.

%FIGURE 1
%\begin{figure} [hbt]
\begin{figure}
\includegraphics[width=8 cm]{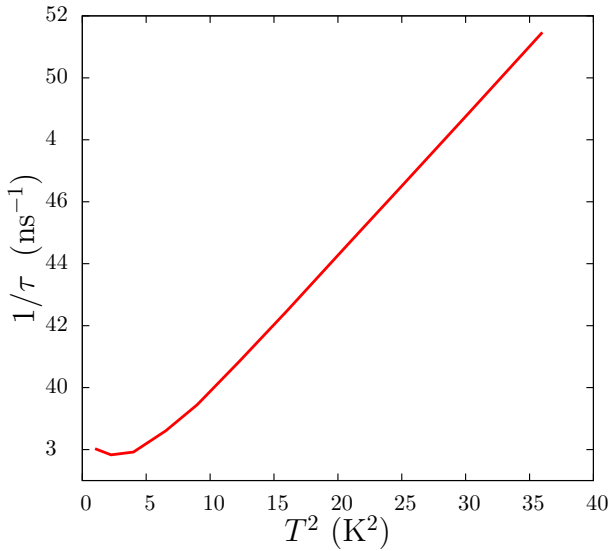}
\caption{
(color online).
Single quantum scattering  rate  $1/\tau_q$ vs $T^2$
}
\label{figure1}
\end{figure}

%FIGURE 2
%\begin{figure} [hbt]
\begin{figure}
\includegraphics[width=8 cm]{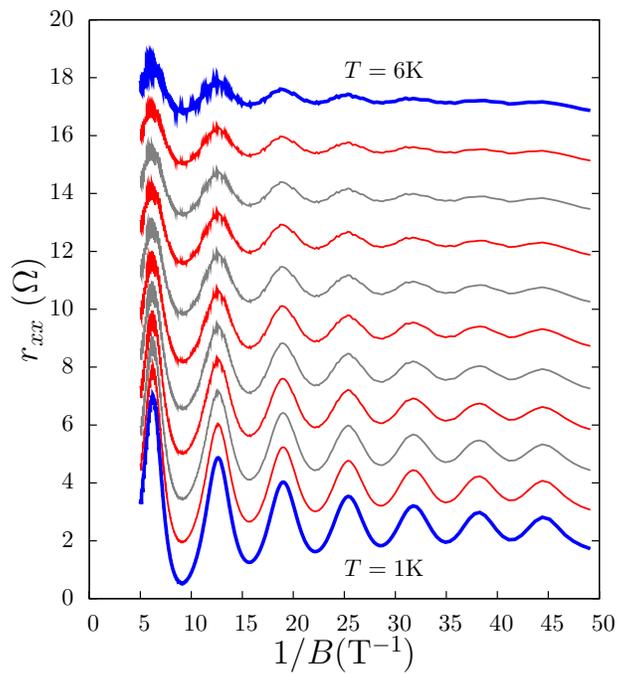}
\caption{
(color online).
Longitudinal differential resistivity $r_{xx}$ as a function
of the inverse of the magnetic field $1/B$ for temperatures
from $T=1 {\rm K}$ (thick [blue] line) to $T=6 {\rm K}$ (thick [blue] line)
in steps of $0.5 {\rm K}$ (thin [gray and red] lines).
}
\label{figure2}
\end{figure}

%FIGURE 3
%\begin{figure} [hbt]
\begin{figure}
\includegraphics[width=8 cm]{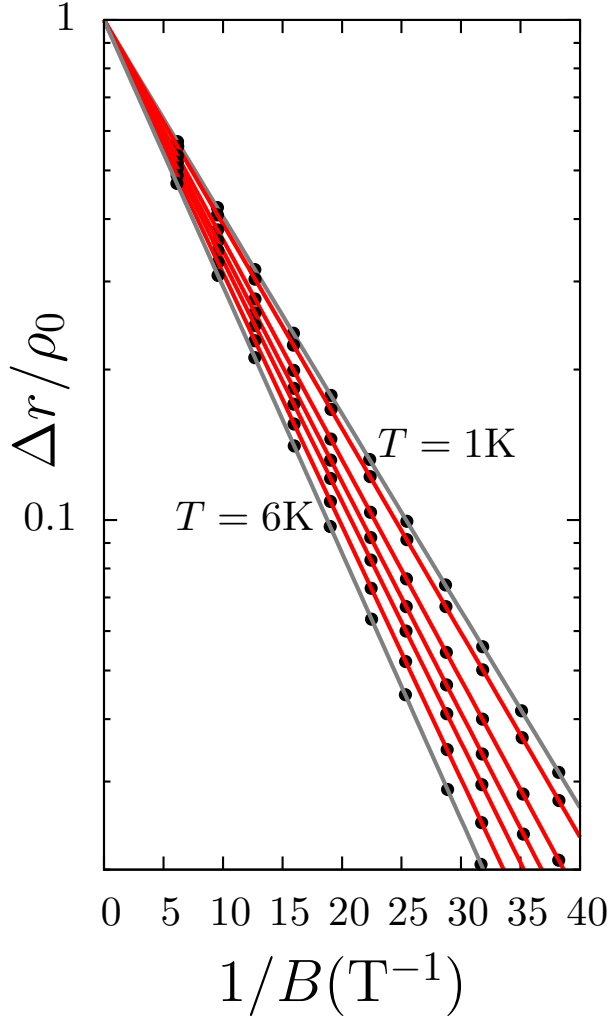}
\caption{
(color online).
Normalized amplitude of the HIRO oscillations
$\Delta r/\rho_0$ as a function of the inverse magnetic field
$1/B$ for the temperatures $T=1,3,4,4.5,5,5.5,6 {\rm K}$ 
for $\alpha_i=25$. The points are obtained from the oscillation  maxima of the plots in Fig. 2,, whereas the  lines  are the  predicted  linear behavior   of the Dingle factor, Eq.(\ref{dingle}).
}   
\label{figure3}
\end{figure}

%FIGURE 4
\begin{figure} [hbt]
%\begin{figure}
\includegraphics[width=8 cm]{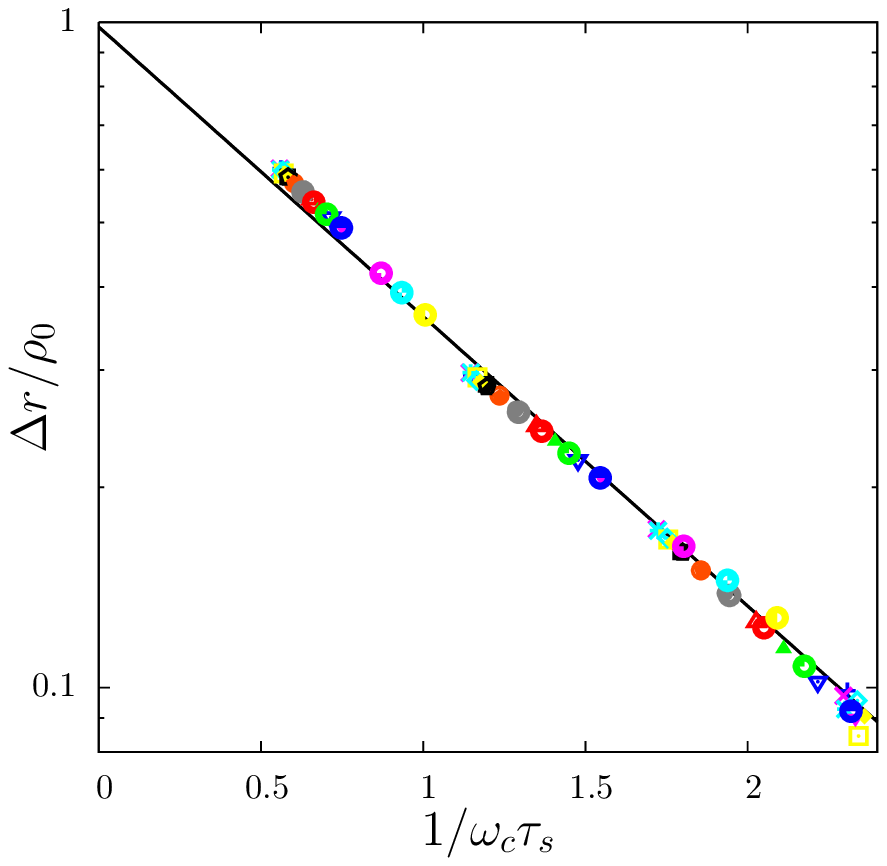}
\caption{
(color online).
 The normalized HIRO amplitude (normalized to the offset of the $\rho_0$)  $\Delta r/\rho_0$
 as a function of $1/\omega_c\tau_s$.
The circles  indicate $\alpha_i=15$ and $\alpha_i=25$, temperatures
ranging from $1$K to $6$K in $0.5$K intervals and magnetic fields between
$0.02$ and $0.2$T. In this plot all data points obtained from our model for different
temperatures, magnetic field and $\alpha$ values fall on the
same universal line (solid line) prescribed by Eq. (\ref{fenomen}). }
\label{figure4}
\end{figure}

Recently there has been considerable interest in the origin of the temperature
dependence of MIRO \cite{studenikin:165321,studenikin:245313,hatke:066804}
and HIRO \cite{hatke:1081}. In both cases the oscillations are best observed
at $T\approx 1 \, K$ and are smeared   when $T$ is of the order of  a few Kelvin.
These experiments show that the amplitude of the oscillations scale
 as   $\exp\left(-a T^2/B\right)$. 
In this paper we present a model that correctly reproduces
the temperature dependence of HIRO.
Both the effects of elastic impurity scattering
as well as those related to inelastic processes due to
electron-electron scattering play an important role.
The model is based on the exact solution of the von Neumann equation
for a 2D damped electron gas subjected to arbitrarily strong magnetic
and dc electric fields.
In addition the model incorporates the weak effects of randomly distributed
neutral and charged impurities through second order time dependent perturbation theory.
This procedures yields a Kubo-like formula  for the electric density
current that incorporates the non-linear dependence on the  electric field.
Both inter- and intra-Landau level transitions contribute to the density current.

\section{Model}

Our starting point is the Hamiltonian for an electron in the
effective mass approximation in two dimensions,  subject to a uniform
perpendicular magnetic field $\boldsymbol B=B\boldsymbol{k}$,
an in-plane electric field
$\boldsymbol E=E_x\boldsymbol{i}+E_y\boldsymbol{j}$, and the  impurity scattering 
potential $V\left(\boldsymbol{r}\right)$. Therefore the electron dynamics
is governed by the total Hamiltonian 
$H = H_e + V$, with
\begin{equation}\label{ham0}
H_e=H_0
+e\boldsymbol{E}\cdot \boldsymbol x \, , 
\end{equation}
here $H_0=\boldsymbol{\Pi}^2/2m$, $m$ is the effective   electron mass, $e$
is the electron's charge,
$\boldsymbol{\Pi}=\boldsymbol{p}+e\boldsymbol{A}$ is the velocity operator and the
vector potential   for an homogeneous magnetic field in the symmetric gauge is given as
$\boldsymbol{A}=\left(-By,Bx\right)/2$.
We consider the impurity scattering potential to be composed of  a combination of
short-range uncorrelated neutral delta scatterers and charged scatterers
\begin{eqnarray}
V\left(\boldsymbol{r}\right)=
{\rm e}^{-\Gamma\left\vert t\right\vert}
\sum_i^{N_{n}}\int\frac{d^2q}{\left(2\pi\right)^2}
V_n\left(q\right)
\exp\left[i\boldsymbol{q}\cdot\left(\boldsymbol{r}
-\boldsymbol{r}_{i}\right)\right]\nonumber\\
+{\rm e}^{-\Gamma\left\vert t\right\vert}
\sum_j^{N_{c}}\int\frac{d^2q}{\left(2\pi\right)^2}
V_c\left(q\right)
\exp\left[i\boldsymbol{q}\cdot\left(\boldsymbol{r}
-\boldsymbol{r}_{j}\right)\right]\label{imppot},
\end{eqnarray}
where $\boldsymbol{r}_{i}$ is  the position of the $i$th neutral scatterer,
$\boldsymbol{r}_{j}$ is the position of the  $j$th charged impurity and
$N_{n}$ and $N_{c}$ are the number of neutral and charged impurities.
The explicit form of the Fourier coefficients
$V_n\left(q\right)$ and $V_c\left(q\right)$ depends
on the nature of the scatterers.
Neutral scatterers are characterized by a
constant\cite{davies:1998,torres:115313,kunold:3803}
\begin{equation}\label{neutral}
V_n\left(q\right)=\frac{2\pi \hbar^2\alpha}{m},
\end{equation}
where $\alpha$ is a dimensionless parameter. The explicit calculations only allows us to calculate the product $\alpha^2 N_n$, hence, without lost of generality, it is  set as $\alpha \equiv 1$.
The potential for charged impurities is modified by screening effects
and takes the form
\begin{equation}\label{charged}
V_c\left(q\right)=\frac{\pi \hbar^2}{m}\frac{e^{-qd}}{1+\frac{q}{q_{TF}}},
\end{equation}
where $d$ is the thickness of the doped layer, $\epsilon_b$ is
the relative permittivity of the surrounding media,    
   and the Thomas-Fermi wave number is given by
\begin{equation}
q_{TF}=\frac{e^2 m}{2\pi\epsilon_0\epsilon_b\hbar^2}.
\end{equation}
The adiabatic switching of the impurity potential at the initial time $t_0\to -\infty$
is controlled by  the factor $\exp\left(-\Gamma\vert t\vert\right)$.

Now we turn to the calculation of the density current.
It is computed as the time and thermal average of the velocity operator
\begin{eqnarray}\label{denscurr}
\boldsymbol{J}=\,    \left[ \frac{e}{T \, S}\int_{- T/2}^{T/2}dt{\rm Tr}
\left[\rho\left(t\right) \boldsymbol{\Pi}\right]  \right]_{T \to \infty}  ,
\end{eqnarray}
where $S$ is the  sample surface,  and $\rho\left(t\right)$
is the density matrix operator    obtained from  the solution  of the
 von Neumman's  equation  $i\hbar\partial \rho/\partial t=\left[H,\rho\right]$.
This  equation is  solved by means of a series of
unitary  transformations  \cite{kunold:205314}. The magnetic and electric field
part of the Hamiltonian  are  treated exactly  by means of the first unitary transformations, that is expressed in terms of functions that solve the corresponding classical equations.
Whereas the impurity  potential is
incorporated through second order perturbation theory.
It is assumed that, in the absence of the impurity,
potential the density matrix reduces to the equilibrium density matrix given
by $\rho_0 = f(H_e)$, with $f(E)$ the Fermi
distribution function.

The computation of Eq. (\ref{denscurr}) yields an
explicit expression for
the  current density. It splits into the Drude and an impurity
induced contribution:
\begin{equation}\label{dencu1}
\boldsymbol{J}= \boldsymbol{J}^D + \boldsymbol{J}^{imp} \, .
\end{equation}
The Drude contribution is given by $\boldsymbol{J}^D  = n_e \, e \,\boldsymbol{v}_d$,
where $n_e$ is the electron density and the drift velocity is  given
by
\begin{equation}\label{vderiva}
\boldsymbol{v}_d =  \frac{ e\tau^{in}_{tr}}{m} \,
\frac{\boldsymbol{E }-\omega_c \tau^{in}_{tr} \,
\boldsymbol{k} \times \boldsymbol{E}}
{1+( \omega_c \tau^{in }_{tr} )^2} \, .
\end{equation}
The  transport inelastic scattering time $\tau^{in}_{tr}$ arises from a damping term $\boldsymbol{v} /\tau^{in}_{tr}$ that is  added to the classical equation of motion in order to incorporate dissipative effects. 
The components of impurity induced density current can be expressed as
\begin{eqnarray}\label{dencu2}
\boldsymbol{J}^{im}=
\frac{e \omega_c }{\hbar^2}\sum_{\mu\mu^{\prime}}
\int \frac{d^2q}{2 \pi}
\left[f_{\mu, q/2}-f_{\mu^\prime, -q/2}\right]
\, \left(n_n\boldsymbol{G}^n_{\mu\mu^{\prime}}
+n_c\boldsymbol{G}^c_{\mu\mu^{\prime}}\right)
\end{eqnarray}
where  $n_n = N_n/S $ and $n_c = N_c/S $ are the density of neutral and charged impurities;   $f$ is the Fermi distribution function evaluated at the
tilted LL energies ${\cal E}_{\mu, q}= \hbar \omega_c (\mu + 1/2) + \hbar \omega_q$  where 
$ \omega_q = \boldsymbol{q}  \cdot \boldsymbol{v}_d$.
The function $\boldsymbol{G}_{\mu\mu^{\prime}}$ is given by
\begin{equation}\label{dosexy}
\boldsymbol{G}^{n,c}_{\mu\mu^{\prime}}
=\left\vert V_{n,c}\left(q\right) \right\vert^2
\left\vert D_{\mu\mu^{\prime}}
\left(z_q\right)\right\vert^2 
\frac{\boldsymbol{q} \Delta_{\mu\mu^{\prime}}
+2\boldsymbol{{\tilde q}}\omega_c\Gamma}
{\Delta^2_{\mu\mu^{\prime}}+4\omega_c^2\Gamma^2},
\end{equation}
where  $\boldsymbol{q}=\left(q_x,  q_y\right)$ is the  transfer momentum, and   $\boldsymbol{{\tilde q}}=\left(q_y, - q_x\right)$  its dual, whereas
\begin{equation}\label{dosexy2}
\Delta_{\mu\mu^{\prime}}
 =\left[\omega_q+\omega_c\left(\mu-\mu^{\prime}\right)\right]^2
-\omega_c^2+\Gamma^2 \, .
\end{equation}
Finally the  matrix elements $D_{\mu,\nu}$ are given by
\begin{multline}\label{matD}
D_{\mu\mu^{\prime}} \left(z_q\right) =
\exp\left(-\frac{\left\vert z_q\right\vert^2}{2}\right)
\left\{\begin{array}{ll}
z_q^{\mu-\mu^{\prime}}\sqrt{\frac{\mu^{\prime}!}{\mu!}}
L_{\mu^{\prime}}^{\mu-\mu^{\prime}}\left(\left\vert z_q \right\vert^2\right),
\mu\ge\mu^{\prime},\\
\left(-{z_q}^*\right)^{\mu^{\prime}-\mu}
\sqrt{\frac{\mu!}{\mu^{\prime}!}}
L_{\mu^{\prime}}^{\mu^{\prime}-\mu}\left(\left\vert z_q\right\vert^2\right),
\mu\le\mu^{\prime},
\end{array}\right.
\end{multline}
where  $z_q =l_B(q_x-iq_y)/\sqrt{2}$,  and $L_\nu^{\mu-\nu}$ denotes
the associated Laguerre polynomial.

Though $\boldsymbol{J}^D$ depends linearly on the electric field,
$\boldsymbol{J}^{imp}$
introduces a nonlinear contribution through $\omega_q$.
From the Lorentzian form of the $G_{\mu\mu^\prime}^{n,c}$ function
in Eq. (\ref{dosexy}) it can be seen that the $\Gamma$ parameter effectively
controls the LL broadening\cite{sinova:235203}.
We thus introduce    disorder broadening effects
through $\Gamma$,
moreover, in order to take into account the LL width dependence on the
magnetic field henceforth we shall consider\cite{ando:437,potts:5189}
\begin{equation}\label{width}
\Gamma=\sqrt{\frac{2 \hbar^2\omega_c}{\pi \tau_{q}}} , 
\end{equation}
as expected we assume that the LL width depends on the quantum scattering times $  \tau_{q}$.

The differential conductivity tensor is calculated from
$\sigma_{ij}=\partial J_i/\partial E_j$ and the  differential
resistivity tensor is obtained from the inverse of the conductivity:
that is $r_{ij}=\sigma^{-1}_{ij}$.

HIRO experiments are carried out at a fixed longitudinal current density
$J_{dc}$ and zero transverse current.
The longitudinal electric field $E_x$ and the Hall field $E_y$
are thus computed from the following implicit equations
\begin{eqnarray}\label{condi}
J_x\left(E_x,E_y\right)=J_{dc}, && J_y\left(E_x,Ey\right)=0,
\end{eqnarray}
where $J_x$ and $J_y$ are given by Eqs. (\ref{dencu1}-\ref{dosexy2}).
Given the complexity and
the nonlinear character of the current density expressions,
the solution of Eq. (\ref{dosexy2}) is computed by a recursive  numerical iteration, 
until the accuracy goal $J_y/J_x \approx 1\times 10^{-15}$ is reached
\cite{kunold:3803}.

%%%%%%%%%    Aqui  %%%%%%%%%%%%% 
In the present model the current density Eq. (\ref{dencu1}) depends directly  on the transport inelastic scattering time  Eq. (\ref{vderiva}),
and the quantum scattering time Eq. (\ref{width}). On the other hand, in order to make contact with experimental results we need  to specify the mobility, that is given in terms of the total transport time $\mu = e \tau_{tr}/m$. Using the 
  Matthesiessen's rule it is possible to incorporate both  inelastic and impurity  contributions to the 
transport scattering time as
\begin{equation}\label{mattrule}
\frac{1}{\tau_{tr}}=\frac{1}{\tau^{in}_{tr}}+\frac{1}{\tau^{imp}_{tr}}.
\end{equation}
 
The  transport as well as the quantum  scattering times associated to the
impurity potentials are given by\cite{davies:1998,dittrich:1998}
\begin{eqnarray}\label{tqimp}
\frac{1}{\tau^{imp}_{q}}=\frac{m}{\pi\hbar^3k_F}\left[n_n
\int_0^{2k_F}{dq
\frac{\left\vert V_n\left(q\right)\right\vert^2}
{\sqrt{1-(\frac{q}{2k_F})^2}}}\right.\nonumber\\
+\left.n_c
\int_0^{2k_F}{dq
\frac{\left\vert V_c\left(q\right)\right\vert^2}
{\sqrt{1-(\frac{q}{2k_F})^2}}}\right]\label{taus}
\end{eqnarray}
and
\begin{eqnarray}\label{ttrimp}
\frac{1}{\tau^{imp}_{tr}}=\frac{m}{2 \pi\hbar^3k_F^3}\left[n_n
\int_0^{2k_F}{   q^2 dq 
\frac{\left\vert V_n\left(q\right)\right\vert^2}
{\sqrt{1- (\frac{q}{2k_F})^2}}}\right.\nonumber\\
+\left.n_c
\int_0^{2k_F}{  q^2 dq 
\frac{\left\vert V_c\left(q\right)\right\vert^2}
{\sqrt{1-(\frac{q}{2k_F})^2}}}\right],\label{taut}
\end{eqnarray}
where $n_n$ and $n_c$ are the neutral and charged impurity
densities respectively,  and $k_F$ is the Fermi momentum.
For  the quantum inelastic time we assume that  it  arises from  electron-electron scattering, 
that is known to be well  estimated 
 by the following expression
\cite{giuliani:4421,hu:10072}
\begin{equation}\label{taui}
\frac{1}{\tau^{in}_{q}}=\frac{E_F}{\hbar}\left(\frac{k_BT}{E_F}\right)^2
\left[\ln\left(\frac{k_BT}{E_F}\right)
+\ln\left(2\frac{q_{TF}}{k_F}\right)+1\right].
\end{equation}
For typical experimental conditions, the following condition: 
$\ln\left(\frac{k_BT}{E_F}\right)
+\ln\left(2\frac{q_{TF}}{k_F}\right)\ll 1$  holds;
thus $1/\tau^{in}_{q}$ has mainly a $T^2$ behaviour
\begin{equation}\label{tauiap}
\frac{1}{\tau^{in}_{q}} \approx \frac{E_F}{\hbar}\left(\frac{k_BT}{E_F}\right)^2.
\end{equation} 
We have explicit expressions for the quantum  impurity Eq. (\ref{tqimp}) and quantum inelastic  Eq. (\ref{tauiap}) scattering  times; hence it is straightforward to determine  the total  quantum scattering rate using  

\begin{equation}\label{mattrule2}
\frac{1}{\tau_q}=\frac{1}{\tau^{imp}_q}+\frac{1}{\tau^{in}_{q}}.
\end{equation}
 
The determination of the transport inelastic time is more involved given the complex structure of the
collision integral\cite{hu:10072}. Thus, for simplicity, we shall assume a simple relation relation between the transport and quantum  inelastic scattering times:
$\tau^{in}_{tr}=\alpha_{i} \tau^{in}_{q}$ where $\alpha_{i}$ is a constant. Now we can use Eq.(\ref{mattrule}) to determine the total transport time.

\section{Results and discussion}

We consider a sample of electron surface density and mobility of
$n_e=3.7 \times 10^{11} cm^{-2}$ and $\mu=1\times 10^7 cm^2/V s$
respectively,
a doped layer of $d=20 nm$ with  $ n_c/n_n=300$  and
$ n_n=1.7\times 10^7 cm^{-2}$ that leads to a  ratio between the quantum and transport  impurity
scattering times of
$\tau^{imp}_{tr}/\tau^{imp}_{q}\approx 11$ according to
Eqs. (\ref{taus}) and (\ref{taut}). The longitudinal current
$I$ is set to $80\mu$A and for the sample width we assume  $w=100\mu$m; hence $J_{dc} = 0.8 \, A/m$.

We have a  formalism in which  the nonlinear resistivity effects,  as well as the relaxation rates can be consistently calculated, once the neutral and charged scattering potentials have been specified. The   quantum  scattering rate  $1/\tau_q$ 
as a function of $T^2$   is presented in Fig. \ref{figure1}.
As the temperature increases,  we observe  the linear behavior related  to the inelastic scattering contribution Eq.(\ref{tauiap}).
On the other hand, the low temperature limit is determined by the impurity scattering contribution  in Eq.(\ref{tqimp}).

In Fig. \ref{figure2} we present the longitudinal differential resistivity
$r_{xx}$ as a function of the inverse magnetic field for temperature values
from $1$K to $6$K in $0.5$K increments.
We observe clear  differential magnetoresistance oscillations periodic in $1/B$,  we see up to the
6th cyclotron resonance. The HIRO originate from the inter-Landau level transitions induced by impurity scattering and are governed by the ratio of  the  Hall to the  cyclotron  frequencies. We observe that the  oscillation peaks appear near integer values of the  
dimensionless parameter  $\epsilon=\omega_H/\omega_c$. Here $\hbar \omega_H =   e E_H (2 R_c)$ is the energy associated with the Hall voltage drop across the cyclotron diameter $R_c = v_F/\omega_c$ and $E_H=  J_{dc} B/ e n_e$ is the Hall field. 
The oscillatory periodic pattern is clearly temperature independent, however we observe that the oscillation amplitude  gradually decay with increasing temperature, and almost disappear at $T\approx 6$K. To further characterize the HIRO temperature dependance as well as its physical origin,  we present  in Fig. \ref{figure3} the normalized HIRO amplitude
$\Delta r/\rho_0=\left(r_{xx}-r_0\right)/\rho_0$
corresponding  to  the  same  data shown in Fig. (\ref{figure1}) and
arrange it in a semilog plot as a function of the inverse of
the magnetic field for temperatures ranging from $2K$ up to $6K$
in $1K$ steps in. Here $\rho_0$ is Drude resistivity and  $r_0$ is the offset of the HIRO. 
It is interesting to observe that the oscillations amplitude clearly display an exponential decay. One also observe that the 
exponent  grows with increasing temperature as $T^2$ and decreases with the magnetic filed as $1/B$. The   results of our detailed calculation  reproduce the most important 
features of  recent  experiments and is also in good agreement with the approximated expression proposed in reference 
\cite{vavilov:115331,hatke:1081}
\begin{equation}\label{fenomen}
r_{xx}=r_0+\rho_0 \frac{\left(4\delta\right)^2}{\pi}
\frac{\tau_t}{\tau_\pi}\cos\left(2\pi\frac{\omega_H}{\omega_c}\right)
\end{equation}
 where  $\tau_{\pi}$ is the time
describing electron backscattering off  impurities and  $\delta$ is the Dingle factor
 \begin{equation}\label{dingle}
\delta=\exp\left(- \frac{\pi}{\omega_c \tau_q}\right) \, ,
\end{equation}
given in terms of the total quantum scattering time Eq.(\ref{mattrule2}).

Our complete calculation depends on  the total quantum scattering rate through the LL width Eq. (\ref{width}) and also on the transport  inelastic rate that appears on the term Drude  $\mathbf{J}^D = n_e e \mathbf{v}_d$ with the drift velocity given by Eq. (\ref{vderiva}). Consequently the differential resistivity is expected to depend on the temperature, magnetic field and the parameter $\alpha_i$ that relates the transport and quantum  inelastic scattering times:
$\tau^{in}_{tr}=\alpha_{i} \tau^{in}_{q}$ where $\alpha_{i}$. However Eqs. (\ref{fenomen},\ref{dingle}) suggest a simple relation for differential resistance on terms of the Dingle factor. 
In Fig.  \ref{figure4}  we present the results of our model  for  the normalized HIRO amplitude $\Delta r/\rho_0$
normalized to the offset of the $\rho_0$ HIRO oscillations as a function of
$1/\omega_c\tau_q$ where the different symbols (colors)
indicate $\alpha_i=15$ and $\alpha_i=25$, temperatures
ranging from $1$K to $6$K in $0.5$K intervals and magnetic fields between
$0.02$ and $0.2$T. In this plot all data points obtained for different
temperatures, magnetic field and $\alpha_i$ values fall on the
same universal line prescribed by Eqs.  (\ref{fenomen},\ref{dingle}). Hence it is concluded that the HIRO temperature dependance  
originates from electron-electron contribution  to  the quantum scattering rate.

\section{Conclusions}

We have presented a model to explain the HIRO properties of   high mobility 2DES, including  its  temperature dependence.
 The effects of the electric and magnetic 
part of the Hamiltonian were treated exactly,  whereas the impurity potential is included  by  means of a 
second order perturbation calculation. Our model contains the effects of neutral delta scatterers and
charged impurities.
We introduced inelastic  effects through the damping  that  appears in the solution of the  classical equation of motion that leads to the Drude contribution Eq.(\ref{vderiva}). The $T^2$ dependence of $\tau_q^{in}$ is required in order to reproduce the correct  temperature dependence of the HIRO amplitude, and is consistent with 
electron-electron inelastic processes. 

%\acknowledgments
We acknowledge support from  UAM-A CBI project 2232204.
A. Kunold wishes to thank INSA-Toulouse for the use of their cluster.

%\bibliography{kunold}

\end{document}